\documentclass[aps,twocolumn,pra,preprintnumbers,amsmath,amssymb,superscriptaddress]{revtex4}
\pdfoutput=1
\usepackage{graphicx}
\usepackage{color}
\usepackage{amsmath,amsfonts,amssymb,bm}
\newcommand{\be}{\begin{align}}
\newcommand{\ee}{\end{align}}
\newcommand{\p}{\partial}

\begin{document}

\title{Cosmic Censorship at Large D: \\Stability analysis in polarized AdS
black branes~(holes)}
\preprint{
{\normalsize OU-HET-958}
}


\author{Norihiro {\sc Iizuka}}\email[]{iizuka@phys.sci.osaka-u.ac.jp}
\affiliation{%
{\it Department of Physics, Osaka University, Toyonaka, Osaka 560-0043, JAPAN 
}}

\author{Akihiro {\sc Ishibashi}}\email[]{akihiro@phys.kindai.ac.jp}
\affiliation{%
{\it Department of Physics, Kindai University, Higashi-Osaka 577-8502, JAPAN
}}

\author{Kengo {\sc Maeda}}\email[]{maeda302@sic.shibaura-it.ac.jp}
\affiliation{%
{\it Faculty of Engineering,
Shibaura Institute of Technology, Saitama 330-8570, JAPAN}}

\begin{abstract}
We test the cosmic censorship conjecture for a class of polarized AdS black branes~(holes) in the Einstein-Maxwell theory at large number of dimensions $D$. We first derive a new set of effective equations describing the dynamics of the polarized black branes~(holes) to leading order in the $1/D$ expansion. In the case of black branes, we construct `mushroom-type' static solutions from the effective equations, where a spherical horizon is connected with an asymptotic planar horizon through a `neck' which is locally black-string shape. We argue that this neck part (of black string) cannot be pinched off dynamically from the perspective of thermodynamical stability. In the case of black holes, we show that the equatorial plane on the spherical horizon cannot be sufficiently squashed unless the specific heat is positive. We also discuss that the solutions are stable against linear perturbation, agreeing with the thermodynamical argument. These results suggest that Gregory-Laflamme type instability does not occur at the neck, in favor of the cosmic censorship. 
\end{abstract}

\maketitle

\section{Introduction}
\label{sec:I}

In contrast to asymptotically flat spacetimes, there is a large variety of asymptotically Anti de Sitter~(AdS) black hole solutions 
due to the warping factor.  For instance, given an asymptotically AdS charged black hole, one can deform it by applying a 
non-uniform electric field and thereby construct a new black hole solution without destroying the asymptotic AdS structure, 
as demonstrated by~\cite{MaedaOkamuraKoga}. 
This fact leads to the recent discovery of four-dimensional polarized AdS black holes in a 
dipolar electric field~\cite{CGOPS2016}. The solution was numerically constructed as a generalization of 
the Ernst solution~\cite{Ernst}.  
Such polarized AdS black holes were extended into four-dimensional polarized AdS black brane 
solutions with a planar horizon, where chemical potential varies along one spatial direction~\cite{HSW2016}.  
In the latter black brane case, by locally applying sufficiently large enough and localized chemical potential,  
it is shown numerically that the configuration of the horizon looks like 
a mushroom and hence is called a ``black mushroom'' solution~\cite{HSW2016}.

In this black mushroom solution, a neck connecting a localized spherical black hole and asymptotic planar horizon appears.  
The neck of the black mushroom solution is getting thiner as the temperature is lowered, and is expected to behave 
like a thin black string. This immediately leads us to the question of whether thin neck part is pinched off dynamically 
due to the Gregory-Laflamme instability~\cite{GL}. If that is the case, a naked singularity would appear and the cosmic 
censorship~\cite{Penrose} would be violated in the polarized black brane. 
There have also been a number of numerical study for the violation of the cosmic censorship in higher dimensions.  
The goal of this paper is to supply an {\it analytic} study for the cosmic censorship conjecture 
in the class of polarized AdS black branes~(holes)  
in the Einstein-Maxwell theory, namely, mushroom-type and other type of black branes~(holes).

In order to attack this issue {\it analytically}, we adopt the {\em large D} effective theory approach 
developed in Ref.~\cite{EST2013, {Emparan:2013xia}, {Emparan:2014cia}, {Emparan:2014jca}, {Emparan:2014aba}, {Emparan:2015rva}, 
{Emparan:2015, Bhattacharyya:2015}}. 
We first derive a tractable set of effective $1+1$-dimensional 
equations describing the dynamics of deformed charged 
AdS black branes~(holes) in the leading order of the $1/D$ expansion. 
Then using these effective equations, 
as in Ref.~\cite{HSW2016}, we obtain polarized black mushroom solutions with a neck connecting a localized spherical 
horizon and an asymptotic planar horizon. Near the neck, the horizon geometry locally behaves as a black string, and 
it is polarized by strong electric field along the neck. 
Applying the claim of the Gubser-Mitra conjecture~\cite{GubserMitra2000, GubserMitra2001}, which has now been proven for some cases~\cite{HW13}, to the local black string, we argue that 
the neck should be locally stable against physically reasonable perturbations, conforming to 
the thermodynamic stability.

We also find polarized AdS black {\em hole} solutions in which the spherical horizon is squashed around 
the equatorial surface. One may expect that there could be a black ``dumbbell'' solution whose 
horizon looks like a dumbbell having two spherical horizons connected by a portion of a thin black string. 
We show however that the equatorial plane on the spherical horizon cannot be sufficiently squashed 
while keeping its local specific heat negative to lead an instability. This implies that  
there is no black dumbbell solution,
where two spherical horizons are connected through a thermodynamically {\it unstable} thin black-string-shape neck, and therefore 
the neck cannot be pinched off dynamically due to the Gregory-Laflamme instability~\cite{GL}. 
We also discuss that the solutions are stable against linear perturbations, being consistent with the thermodynamical argument.

The organization of this paper is as follows;
In the next section, we first derive the $1+1$-dimensional effective equations by expanding the Einstein equations in the inverse power of $D$. 
In sections \ref{sec:III} and \ref{sec:IV}, we construct the black mushroom solutions and test the cosmic censorship conjecture in them. 
In section \ref{sec:V}, we repeat the analysis in the polarized AdS black hole solutions. 
Section \ref{sec:VI} is devoted to summary and discussions.     

\section{Effective equations in Charged AdS black branes}
\label{sec:II}
We start with the following $D$-dimensional Einstein-Maxwell equations with a negative cosmological constant 
\begin{align}
\label{Eq:Ein}
& R_{\mu\nu}-\frac{1}{2}R\,g_{\mu\nu}+\Lambda\,g_{\mu\nu}=\frac{1}{2}\left(F_{\mu\rho}{F_\nu}^\rho-\frac{1}{4} F^2 g_{\mu\nu}    \right) \,, 
\nonumber \\
& \Lambda=-\frac{(D-1)(D-2)}{2L^2} \,, \quad 
\frac{1}{\sqrt{-g}}\partial_\mu \left( \sqrt{-g} F^{\mu\nu} \right) 
=0,  
\end{align}
where $L$ is the AdS curvature length and $F_{\mu\nu}=\p_\mu A_\nu-\p_\nu A_\mu$. 
We make the following ansatz for the metric and the gauge field as
\begin{align}
\label{metric_ansatz_br}
ds^2&=-Adt^2+2u_t\,dtdr-2C_z\,dtdz+G_{zz}dz^2+\frac{r^2z^2}{L^2}\,d\Omega_{n-2}^2, \nonumber \\
A&=A_tdt+A_zdz, 
\end{align}
where $d\Omega_{n-2}^2$ is the metric of unit sphere with $n=D-1$. Note that we do not 
rescale $z$-coordinate, as done in \cite{EST2015} since $z$ is not the direction of Killing symmetry 
of the background geometry.  

For simplicity, we assume that at large $D$, the gauge field $A_\mu$ behaves as 
\begin{align} 
\label{gauge_ansatz}
A_t=O(n^{-\frac{1}{2}}) \,, \qquad A_z=O(n^{-\frac{3}{2}}). 
\end{align}
Then, the electric charge of the black brane can be dealt with a test charge so that it does not affect 
the metric at leading order in the expansion in the inverse of $n$~\cite{EILST2016}.  
This leads to the metric expansion as follows 
\begin{align}
\label{ansatz_II}
 A(r,t,z)&=\frac{r^2}{L^2}\left(1-\frac{m(t,z)}{r^n}\right) \nonumber \\
&\quad +\frac{r^2}{nL^2}\left(\frac{Q(t,z)}{r^{2n-2}}+a_1(r,t,z)\right)+O(n^{-2}), \nonumber \\
 C_z(r,t,z)&= \frac{p(t,z)}{nr^n}+O(n^{-2}), \nonumber \\
 u_t(r,t,z)&=1+\frac{\beta_t(r,t,z)}{n}+O(n^{-2}), \nonumber \\
 G_{zz}&=\frac{r^2}{L^2}+\frac{H(r,t,z)}{n}+O(n^{-2}), 
\end{align}
where the horizon is determined by $A=0$. 
It is convenient to use the formula~(\ref{decomposition}) to expand the Einstein Eqs.~(\ref{Eq:Ein}) order by order 
as a series in $1/n$. Then, we find that the metric given above already solves the Einstein equations at leading order. 

We would like to derive the effective equations 
for the variables, $m(t, z)$, $p(t, z)$, $\cdots$.  
For that purpose, 
let us define $R$ as $R =\left( r/r_0 \right)^n$, 
where $r_0$ is a fiducial horizon size. Hereafter, without loss of generality, we set $r_0=1$. 
We take the large $D$ (or equivalently large $n$) limit in such a way that $R$ 
= {\it fixed}, {\it i.e.,} $r \to1$, $n\to \infty$ with $r^n$ = fixed. 
Note that this limit forces us to set the {\it finite} power of $r$ to be $1$ in the leading order of large $n$ expansion.   
Within this limit, we evaluate the Einstein $\&$ Maxwell equations in the $1/n$ expansion at the horizon and  
derive the effective equation for the variables. 
Note that the horizon is determined as $R = r^n = m(t, z)$ in the leading order of large $n$ expansion.     

With this double scaling limit in our mind, as for the gauge field we make the ansatz for $A_t$ as 
\begin{align}
\label{sol_gauge_t}
& A_t(r,t,z)=\sqrt{\frac{2}{n}}\left(P(t, z) - \frac{q(t, z)}{r^{n }}\right) . 
\end{align}
Here, $P$ plays a role of the chemical potential on the AdS boundary, $r=\infty$ and 
$\p_zP$ corresponds to the external electric field along $z$-direction. Then the $t$-component of 
Maxwell equation at the leading order is automatically satisfied.  The function $q$ will correspond to the electric charge 
as we will see below. 

Substituting Eqs.~(\ref{ansatz_II}) and (\ref{sol_gauge_t}) into the $z$-component of the Maxwell 
equations in Eqs.~(\ref{Eq:Ein}), and by evaluating its leading order at the horizon in the $1/n$ expansion, 
we obtain 
\begin{align}
\label{Maxwell_z}
- \frac{1}{\sqrt{2}}\,\frac{\p A}{\p r}\,\frac{\p A_z}{\p r} - \sqrt{n} \frac{pq}{r^{2n}}+\sqrt{n} \frac{\p P}{\p z} = 0 \,.
\end{align}
This gives a solution for $A_z$ as 
\begin{align}
\label{sol_gauge_z}
A_z(r,t,z)=
\sqrt{\frac{2}{n}}\frac{L^2}{n}\left(\frac{\p P(t,z)}{\p z}\ln(r^n) + \frac{p(t,z)q(t,z)}{m(t,z)r^{n}}\right) \,. 
\end{align}

At next to leading order in $1/n$ expansion, from the $rr$, $rz$, $zz$-component of the Einstein equations, we find that $\beta_t$ and $H$ can be set to zero:  
\begin{align}
\label{gauge_con}
\beta_t=H=0 \,.
\end{align}
From the several components of the Einstein equations~(\ref{Eq:Ein}), we obtain 
\begin{align}
Q=L^2q^2(t,z) . 
\end{align}
Then, $rt$-component of the Einstein equations~(\ref{Eq:Ein})  
reduces to 
\begin{align}
\frac{\p a_1}{\p r}+\frac{1}{n}\frac{\p^2 a_1}{\p r^2}=\frac{nL^4p}{Rz}, 
\end{align} 
and its solution is given by 
\begin{align}
\label{sol_a1}
a_1=-\frac{L^4p\ln R}{zR} \,.
\end{align}
From the $tt$ and $tz$-component of the Einstein equation, the evolution equations for $m$ and $p$ 
are obtained as 
\begin{align}
\label{Eq_mass}
& \qquad \qquad \quad 
 \frac{\p m}{\p t}+\frac{L^2}{z}p-\frac{L^2}{z}\frac{\p m}{\p z}=0 \,,   \\
\label{Eq_pressure}
&\frac{\p p}{\p t}+\frac{L^2}{mz}p^2 - 2q\frac{\p P}{\p z}+\frac{1}{L^2}\frac{\p m}{\p z}
-L^2\frac{\p }{\p z}\left(\frac{p}{z}  \right)=0 \,,
\end{align}
on the horizon $R= r^n=m$. Finally the $r$-component of the Maxwell equations yields the evolution 
equation for $q$: 
\begin{align}
\label{eq_q}
\frac{\p q}{\p t}-\frac{L^2}{z}\frac{\p q}{\p z} + \frac{L^2m}{z}\frac{\p P}{\p z}+\frac{L^2p}{zm}q=0 \,.
\end{align} 
These three equations~(\ref{Eq_mass}), (\ref{Eq_pressure}), and (\ref{eq_q}) are the $1+1$-dimensional effective 
equations for the charged black brane. As far as we are aware, these are completely new equations.     
\section{Black mushroom solutions}
\label{sec:III}
In this section, we derive a black mushroom solution from our effective 
equations (\ref{Eq_mass}), (\ref{Eq_pressure}), and (\ref{eq_q}). The topology of the 
horizon at $t=\mbox{constant}$ surface is $\bm{\mbox{R}}^{n-1}$ and the metric 
becomes 
\begin{align}
 ds^2_{\mbox{\tiny{\it{fixed \hspace{-0.4mm}$t$}}}} =\frac{r^2(z)}{L^2}(dz^2+z^2d\Omega_{n-2}^2) \,, 
\end{align} 
where $z$ 
is the radial coordinate on the horizon and $r(z)$ is the location.  
In the black mushroom solution there is a neck, as found in Ref.~\cite{HSW2016}. By denoting 
the area of the $z=\mbox{constant}.$ surface as $S(z)$, the minimum condition for the 
existence of a neck~(located at $z=z_0~(>0)$) can be geometrically defined as 
\begin{align}
\label{neck_cond}
& \frac{\p S(z)}{\p z}\biggl{|}_{z=z_0}=0, \quad 
\frac{\p^2 S(z)}{\p z^2}\biggl{|}_{z=z_0}>0, \quad m(z_0)<m|_{(z \rightarrow \infty)}, \nonumber \\
& S(z):=\frac{C_0}{L^n}\,R(z)z^n=\frac{C_0}{L^n}\,m(z)z^n,  
\end{align} 
where $C_0$ is the surface area of unit $n-2$-dimensional sphere. The third condition on the 
first line implies that there should be a concavity for the horizon radius $R_H$ in the range $0<z<\infty$. 
When $m=\mbox{constant}.$, 
the solution becomes a plane-symmetric charged black brane solution with no neck. 
As is shown below, the neck can be created only by the localized chemical potential, $P(z)$, or it 
would be more correct to say that the neck can be created by the strong external electric field, $\p_z P$. 

Hereafter, we set $L=1$ for simplicity. Making the ansatz
\begin{align}
m=m(z), \quad p = p(z), \quad q=q(z), \quad P=P(z) \,,
\end{align}
for the static solution, we reduce Eqs.~(\ref{Eq_mass}),  (\ref{eq_q}), and (\ref{Eq_pressure}) to  
\begin{align}
\label{static_effective_brane}
 m' &=p  \,, \nonumber \\
 q' &=\frac{pq}{m}  + mP' \,, \nonumber \\
\left(\frac{p}{z} \right)' &=m'+\frac{p^2}{mz}- 2qP' \,, 
\end{align}
where the prime means the derivative with respect to $z$. 
The horizon is determined by $A=0$ as 
\begin{align}
\label{horizon_hole}
R_H=m-\frac{a_1m}{n}-\frac{Q}{nm}+O\left(\frac{1}{n^2}\right). 
\end{align}
The surface gravity $\kappa$ is given by 
\begin{align}
\label{surfacegravity}
\kappa:=\frac{1}{2}\frac{\p A}{\p r}\biggl{|}_{R=R_H}=\frac{1}{2}\left(n+\ln m-\frac{p}{zm}-\frac{q^2}{m^2}\right)+O\left(\frac{1}{n}\right) .
\end{align}
In order to derive this, one has to be careful to the fact that 
\begin{align}
\label{radial_deformation}
r = 1 + \frac{1}{n} \ln m + O\left(\frac{1}{n^2} \right) .
\end{align}
It is easily checked from Eqs.~(\ref{static_effective_brane}) that $\kappa$ is constant along the horizon by showing that 
\begin{align}
\frac{\p \kappa}{\p z}=O\left(\frac{1}{n} \right).  
\end{align} 
Note that Eq.~(\ref{radial_deformation}) indicates that deformation from the homogeneous black brane 
solution is $O(1/n)$ in our black mushroom solution, while it is $O(1)$ in the black mushroom solution numerically 
constructed in Ref.~\cite{HSW2016}. Nevertheless, as we will show, our black mushroom solution has a neck 
defined in Eq.~(\ref{neck_cond}).  

Now, we will construct a black mushroom solution which is deformed by the external 
electric field $P'$. Substitution of $p=m'$ into the second equation in (\ref{static_effective_brane}) 
yields  
\begin{align}
\frac{q'}{m}-\frac{m'}{m^2}\,q= P'. 
\end{align}
This can be integrated as 
\begin{align}
\label{sol_P}
\frac{q}{m}=P+C,  
\end{align}
where $C$ is an integral constant.  
As an asymptotic boundary condition at infinity, $z \rightarrow \infty$ on the horizon, we will impose 
that the black brane solution asymptotically approaches a uniformly charged black brane solution.
This is equivalent to impose the following conditions, 
\begin{align}
\label{asy_condition}
\lim_{z\to \infty}P &=\frac{q_0}{m_0}, \quad \lim_{z\to \infty}m=m_0~(>0), \nonumber \\
\lim_{z\to \infty}q &= \, q_0~(>0) \,.
\end{align}
The first condition implies that there is no external electric field at infinity. 
The boundary condition determines the integral constant $C$ as 
\begin{align}
\label{constant_C}
C= 0  \,.
\end{align}
This is consistent with the regularity condition on the horizon, $A_t=0$ in 
Eq.~(\ref{sol_gauge_t})~(see, for example, Ref.~\cite{Gubser}). 

Introducing new variables $M$ and $\xi$ as 
\begin{align}
\label{variable_M}
m=m_0\,e^M, \qquad q=\xi\,e^M, 
\end{align}
we obtain the equation of motion for $M$ from the third equation in (\ref{static_effective_brane}) 
as
\begin{align}
\label{Eq_M_br_second}
\frac{M''}{z}-\left(1+\frac{1}{z^2}\right)M'=-\frac{2\xi\xi'}{m_0^2}, 
\end{align}
where we used $p = m' $ and 
\begin{align}
\label{q_P_relation}
P = \frac{q}{m} = \frac{\xi}{m_0}
\end{align}
from Eqs.~(\ref{sol_P}) and (\ref{constant_C}). 
Taking into account that $M\to 0$, $\xi\to q_0$ at $z=\infty$, Eq.~(\ref{Eq_M_br_second}) is integrated as 
\begin{align}
\label{Eq_M}
\frac{M'}{z}-M=\frac{q_0^2-\xi^2}{m_0^2}. 
\end{align}

If there is a neck which is satisfying the conditions~(\ref{neck_cond}) and (\ref{asy_condition}), 
$M$ must have a minimum $M_m$ at 
$z=z_m~(0<z_m<\infty)$~\footnote{If there was no minimum, $M$ would be monotonically decreasing function 
satisfying $M>0$ for $z\in [0,\,\infty)$. This contradicts $m(z_0)<m_0$.}. 
The lower bound of the minimum $M_m$ is determined by Eq.~(\ref{Eq_M}) as
\begin{align}
\label{lower_bound}
M_m\ge -\frac{q_0^2}{m_0^2} = - P^2 |_{z\to \infty}
\end{align}
where the equality is satisfied only when $\xi(z_m)=0$. 
This implies that the {\it minimal} horizon radius $R_H$ around the neck is 
determined by the asymptotic value of 
the chemical potential $P$ given by Eq.~(\ref{q_P_relation}). 

There are infinite degrees of freedom to choose a function $M$ satisfying Eq.~(\ref{Eq_M}), the neck 
condition~(\ref{neck_cond}), and the lower bound~(\ref{lower_bound}). Once we choose a function $M$ satisfying these 
conditions~(\ref{neck_cond}) and (\ref{lower_bound}), $\xi$ and $P$ are 
determined 
from Eqs.~(\ref{Eq_M}) and (\ref{q_P_relation}) 
\footnote{Normally given $P$, the chemical potential on the boundary, the bulk profile $M$ is determined. Here 
we are solving this in a opposite way; given $M$, the bulk profile, we determine the boundary chemical potential profile $P$ which realizes this bulk 
profile $M$.}. 
For example, let us choose a Gaussian like function $M$:     
\begin{align}
\label{sol_M_brane}
M=-B\frac{z^4}{a^4}e^{-\frac{(z-a)^2}{b^2}}, \qquad B>0
\end{align}
to satisfy the asymptotic boundary condition~(\ref{asy_condition}), where $a$, $b$, and $B$ are 
some positive constants. Here, we set $M$ sufficiently rapidly approaches zero at the origin of spherical symmetry, 
$z=0$ to avoid a 
singularity.  
The minimum takes at   
\begin{align}
z_m=\frac{a+\sqrt{a^2+8b^2}}{2}\simeq a+\frac{2b^2}{a} 
\end{align}
in the limit $a\gg b$. To satisfy the lower bound~(\ref{lower_bound}), we choose the parameter $B$ so
that 
\begin{align}
\label{B}
B=B_0\frac{q_0^2}{m_0^2}\left(1+\frac{2b^2}{a^2} \right)^{-2}\simeq  B_0\frac{q_0^2}{m_0^2}, 
\end{align} 
where $B_0$ is a positive constant satisfying $B_0<1$. 
\begin{figure}[ht]
\centering
  \includegraphics[width=\linewidth,clip]{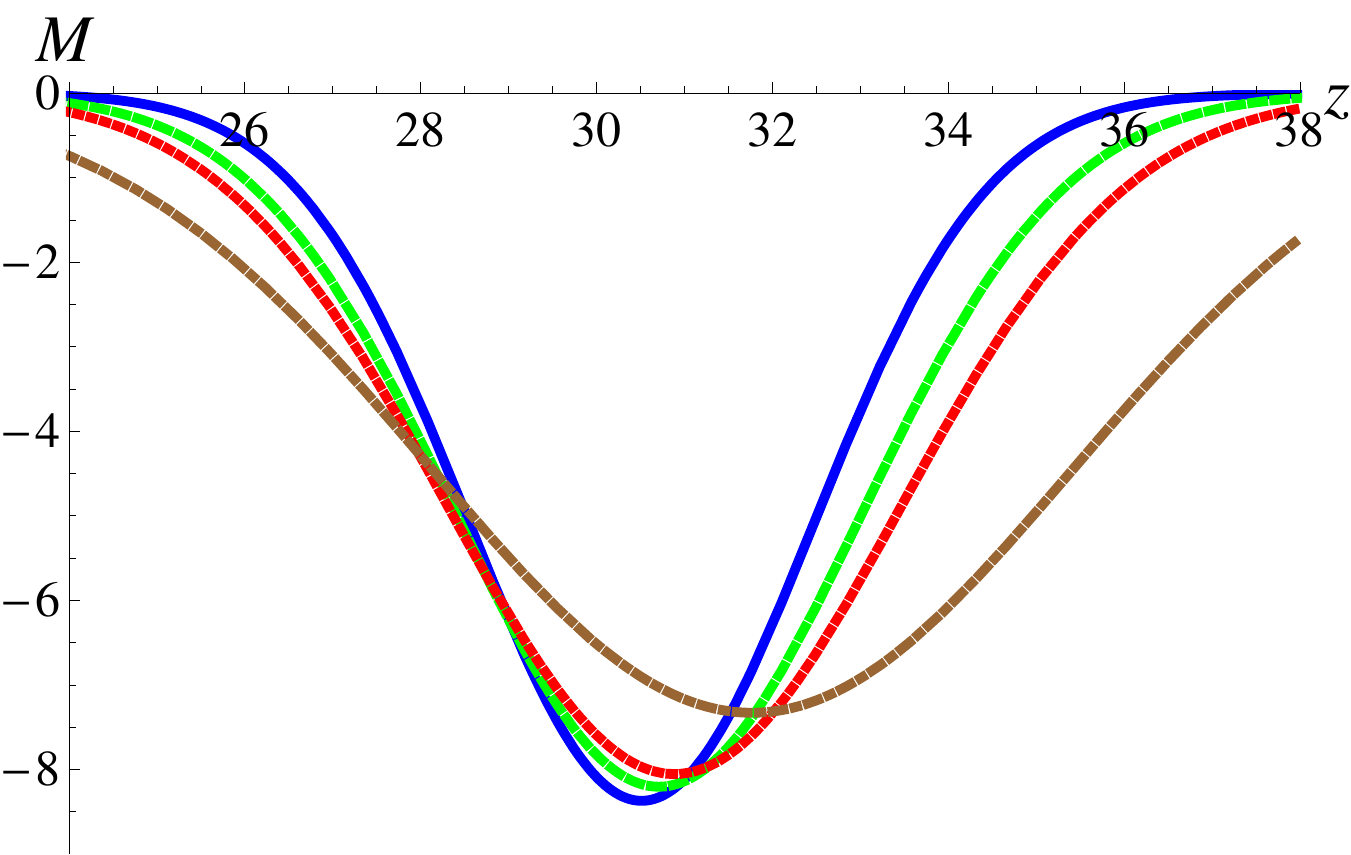} 
\caption{The plot of $M$ for various values of $b=2.8$~(blue, solid), $3.3$~(dashed green), $3.7$~(dotted red), 
and $5.3$~(dotdashed, brown) in the case $a=n=30$, $m_0=1$, $q_0=3$, and $B_0=0.98$}
\end{figure}
Since the horizon is determined by Eq.~(\ref{horizon_hole}), the cross-sectional area $S$ defined 
in Eq.~(\ref{neck_cond}) becomes 
\begin{align}
\label{Sderivativeformula}
S'=S(z)\left(\frac{n}{z}+M'  \right). 
\end{align}
\begin{figure}[ht]
\centering
  \includegraphics[width=\linewidth,clip]{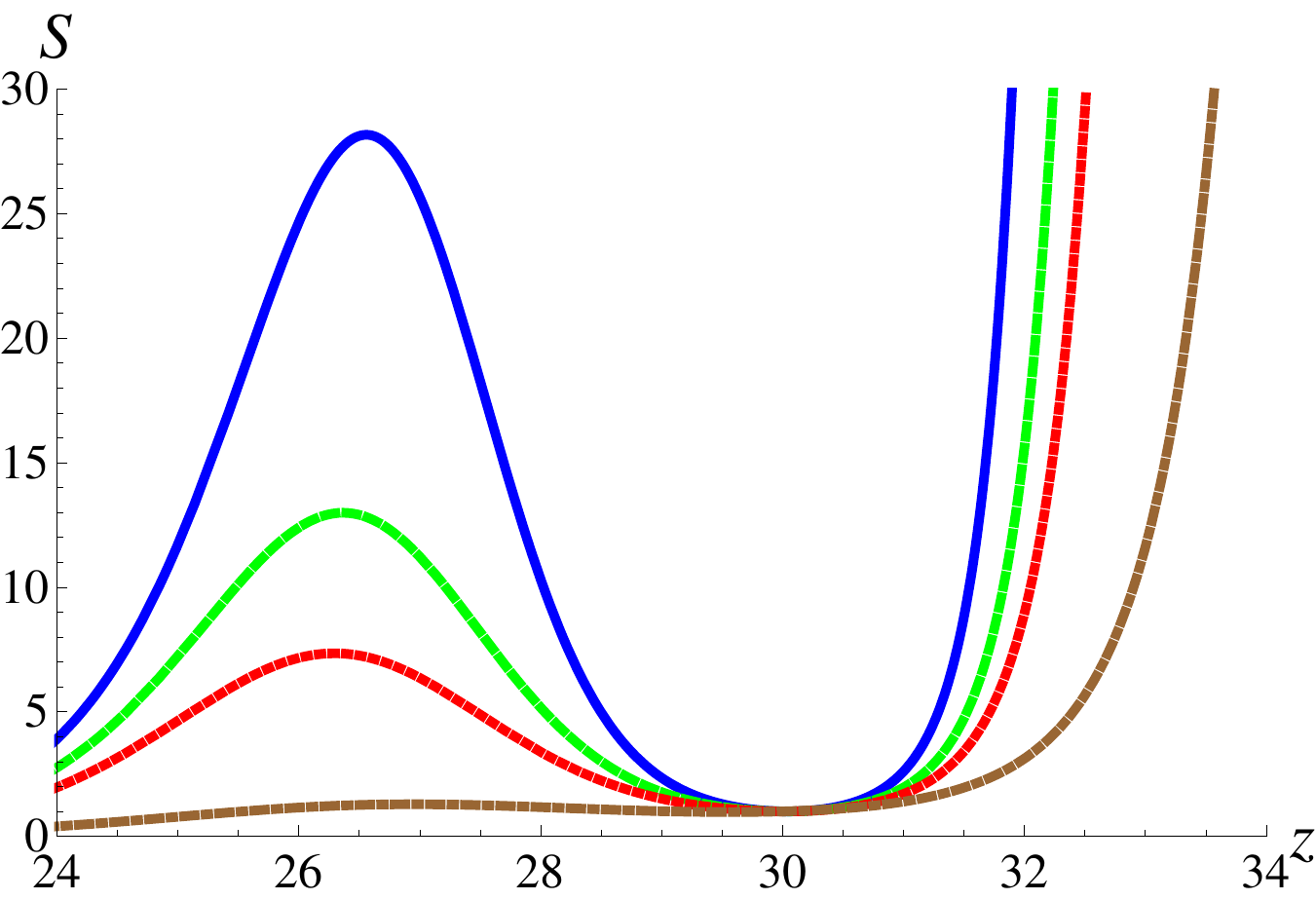} 
\caption{The plot of $S$~(normalized by $S(n)$) for the same values of $b$ as Fig.~1 in the case 
$a=n=30$, $m_0=1$, $q_0=3$, and $B_0=0.98$}
\end{figure}
Note that the expansion (\ref{ansatz_II}) is valid when $M'=O(1)$, therefore setting  
$a=na_0~(a_0>0)$, one obtains   
\begin{align}
M'|_{z = a-b}\simeq -\frac{2B}{b}e^{-1}+O\left(\frac{1}{n} \right) \,.
\end{align}
\begin{figure}[ht]
\centering
  \includegraphics[width=\linewidth,clip]{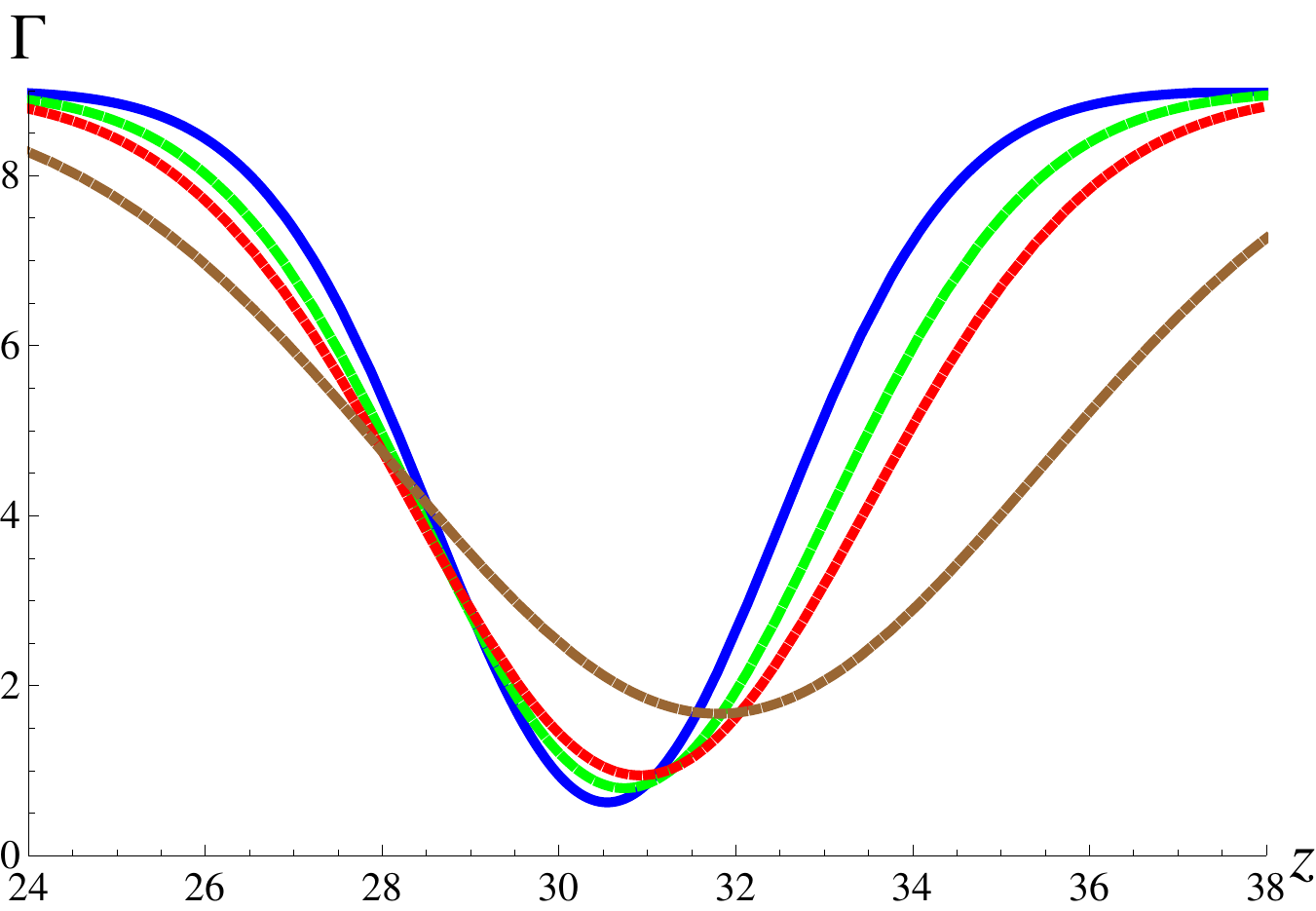} 
\caption{The plot of $\Gamma=P^2$ for the same values of $b$ as Fig.~1 in the case 
$a=n=30$, $m_0=1$, $q_0=3$, and $B_0=0.98$}
\end{figure}
This implies that $S$ must have a minimum around $z=a$ if
\begin{align}
\frac{2a_0B}{b}e^{-1}>1.  
\end{align}

Fig.~1. and Fig.~2. show the plots of $M$ and $S$ near the minimum for various values of $b$ in the case $a=n=30$, 
$m_0=1$, $q_0=3$, and $B_0=0.98$. The cross-sectional area $S$ monotonically increases before reaching the maximum, 
and then 
decreases toward the minimum.  This implies that 
the horizon behaves as a spherical black hole in the region $0\le z<a$, 
and it is connected to an asymptotic planar horizon $z\gg a$ through a neck around $z=a$. To satisfy the 
condition $M'=O(1)$, $a$ must increase as $n$ increases. So, the position of the neck goes away from the center, $z=0$, 
and the neck connects a large spherical black hole with an asymptotic planar horizon, as $n$ increases. 
Note that $S'/S$ increases with the magnitude $O(n)$ before reaching $z = O(n)$, and then decrease with the magnitude $O(1)$ at $z = O(n)$. This implies that the mushroom shape is extremely flattened.  

As shown in Fig.~2, a plateau region appears for each value of $b$, corresponding to the neck in the black mushroom 
solution. This region spreads as $b$ increases, and the spherical black hole portion tends to disappear. 
These facts imply that the black mushroom solution locally approaches a black string solution with translational 
symmetry along $z$ as $b$ increases.  
As shown in Fig.~3, the chemical potential $P$ possesses a precipitous valley near the plateau region. As the 
external electric field $E$ is given by $P'$, the black string portion is supported by the strong electric field. 

\section{Stability analysis}
\label{sec:IV}
As seen in the previous section, we showed that there is a black mushroom solution in which a small spherical black hole 
is connected to the asymptotic planar black brane through a neck that resembles a black string solution. 
In this section, we argue the stability of the black mushroom solution from the perspective of 
thermodynamics, as well as that of the dynamical stability with respect to linear perturbations. 
\subsection{Stability analysis: thermodynamical argument} 
Gubser and Mitra have conjectured that the Gregory-Laflamme instability for black branes with a non-compact translational 
symmetry occurs if and only if they are locally thermodynamically unstable~\cite{GubserMitra2000, GubserMitra2001}. 
This claim was proven~\cite{HW13}. 
This implies that if a black-string-shape neck has a translationally invariant portion larger than the threshold wavelength $\lambda_c$ 
beyond which any longer wavelength perturbations are unstable, it tends to break up under the evolution. 

As shown in Ref.~\cite{KolSorkin2004, AGHKLM2007, EST2015}, higher dimensional black string solution with translational 
symmetry suffers from a Gregory-Laflamme instability for short wavelength perturbations. The threshold 
wavelength $\lambda_c$ is approximately given by 
\begin{align}
\label{threshold}
\lambda_c\sim S(z_m)^{1/(n-2)}\sim \frac{z_m}{\sqrt{n}}. 
\end{align}  
Here, to derive the second approximation, we used the fact that the surface area of unit $n-2$-dimensional 
sphere $C_0$ is given by $C_0\sim n^{-n/2}$~\cite{EST2013}. 
In the $z_m\sim n=30$, $b=5.3$ case in the previous section, $\lambda_c\sim 5.4$, which is comparable to the 
length of the neck, $\sim 6$~(recall that $r\simeq 1$), as seen in Fig.~2. So, one would expect that the 
neck with a translationally invariant portion larger than $\lambda_c$ would be unstable against Gregory-Laflamme 
instability unless it is thermodynamically stable. 

The temperature $T$ for the black mushroom solution corresponds to the surface gravity $\kappa$ in Eq.~(\ref{surfacegravity}). 
As $z\sim z_m\sim n\gg 1$ in the neck, the third term proportional to $p$ becomes irrelevant. Then, $\kappa$ is determined by the local charge $q$ and mass parameters $m$ on the neck. 
Since $\kappa$ increases as 
the mass increases for a fixed charge, it should be thermodynamically stable, implying that the neck should also be 
dynamically stable, according to the conjecture.  

Note that the fact that the existence of the neck forces the specific heat positive is independent of the form of $M$. Given $M$, 
Eq.~\eqref{Sderivativeformula} is generic and in order to have a neck part, we have to have $z_m = O(n)$, since $M' = O(1)$. 
Then, from Eq.~(\ref{surfacegravity}), terms with $p/zm$ becomes $O(\frac{1}{n})$ and we always have a positive specific heat. 
These suggests that in the large $D$, the neck part of the mushroom solution is always stable dynamically.


\subsection{Stability analysis: linear perturbation} 
We consider linear perturbation of the black mushroom solution satisfying Eqs.~(\ref{static_effective_brane}). Here, we 
address the issue whether the linear perturbation has an unstable mode without time dependent external 
force $P$. So, we impose the condition   
\begin{align}
\delta P(t,z)=0. 
\end{align}
Linearizing the evolution equations~(\ref{Eq_mass}), (\ref{Eq_pressure}), and (\ref{eq_q}), we obtain the equations for 
perturbation as 
\begin{align}
\label{per_m_br}
&\dot{\delta m}+\frac{\delta p}{z}-\frac{\delta m'}{z}=0 \,, \\
\label{per_p_br}
&\dot{\delta p}+\frac{2p}{mz}\delta p-\frac{p^2}{m^2z}\delta m- 2P'\delta q+\delta m'-\left(\frac{\delta p}{z}  \right)'=0 \,, \\
\label{per_q_br}
&\dot{\delta q}-\frac{\delta q'}{z} + \frac{\delta m}{z}P'+\frac{q}{zm}\delta p+\frac{p}{zm}\delta q
-\frac{pq}{zm^2}\delta m=0 \,, 
\end{align} 
where a dot and prime denote the derivative with respect to $t$ and $z$, respectively. 
Plugging $\delta p= \delta m' - z\delta {\dot m}$ obtained from Eq.~(\ref{per_m_br}) into Eqs.~(\ref{per_p_br}) and (\ref{per_q_br}), we have  
\begin{align}
\label{eq:m:tz}
& z \delta m''-2z^2  \dot{\delta m}' - \left(1+z^2 + 2z \frac{p}{m} \right) \delta m' 
\nonumber \\
&
\quad+ z^3 \ddot{\delta m} + 2z^2 \frac{p}{m}\dot{\delta m}
 + z \frac{p^2}{m^2} \delta m + 2z^2 P' \delta q =0 \,, \\
%
\label{eq:q:tz}
&  \delta q' - z\dot{\delta q} - \frac{p}{m} \delta q 
   = \frac{q}{m}(\delta m' -z \dot{\delta m}) + \left( P' - \frac{pq}{m^2} \right) \delta m \,. 
\end{align}
Note that when $p=0=P$, the above set of equations reduce to the corresponding perturbation equations for the large $D$ limit of 
the Schwarzschild-AdS black brane solution, which should be stable as it has a positive specific heat.        

It is immediate to see from Eq.~(\ref{eq:m:tz}) that near the center $z=0$, 
the general solution of $\delta m$ behaves in a regular manner as 
\begin{align}
\label{condi:z:0}
 \delta m \simeq C_1 + C_2 z^2  \,,   
\end{align}
with $C_1, C_2$ being some constants independent of the values of $p$ and $P$. Choosing $C_1$ and $C_2$ corresponds to 
specifying a particular boundary condition at the center: for instance, $C_1=0$ corresponds to the Dirichlet boundary 
condition. Actually, which choice of the boundary condition we would take is not relevant to the rest of our argument,  
and thus we leave these constants unspecified. 

It also turns out that Eqs.~(\ref{eq:m:tz}) and (\ref{eq:q:tz}) form a parabolic system. 
To see that, let us change the coordinates $(t,z)$ into $(u :=-t, \: v:= 2t+ z^2)$ so that the above two equations are expressed as  
\begin{align}
\label{eq:m:uv}  
&\left(\partial_u^2 -2 \partial_v - \frac{2}{z}\frac{p}{m} \partial_u + \frac{1}{z^2}\frac{p^2}{m^2} \right) \delta m 
 + 4z\partial_vP \delta q =0 \,, 
{}
\\
\label{eq:qm:uv}  
& \left(z\partial_u -\frac{p}{m} \right) \delta q = \frac{q}{m}\left( z\partial_u -\frac{p}{m}  \right) \delta m +2z \partial_vP \delta m \,,
\end{align} 
with $z$ viewed as the function of $(u,v)$. 


Recalling the conditions (\ref{asy_condition}) at $z\rightarrow \infty$ and also noting $p=m' \rightarrow 0$, we find 
Eq.~(\ref{eq:qm:uv}) to become 
\begin{align}
 \partial_u \delta q \simeq \frac{q_0}{m_0} \partial_u \delta m \,,
\end{align}
and thus we have $\delta q \simeq (q_0/m_0) \delta m$. Eq.~(\ref{eq:m:uv}) then asymptotically takes the form of 
thermal diffusion equation:  
\begin{align}
\label{eq:diffusion}
 (\partial_u^2 -2 \partial_v) \delta m \simeq 0 \,. 
\end{align}  
We naturally impose the following regularity conditions at large $z$:
\begin{align}
\label{condi:z:infty}
\lim_{z \rightarrow \infty} \delta m=0 \,.  
\end{align} 

Provided the separation of variable, the above equation can be immediately solved as 
\begin{align}
\label{expr:delta:m}
\delta m &= \sum_\lambda a(\lambda) e^{-\lambda^2v} \cos(\sqrt{2}\lambda u + \theta_\lambda) 
\nonumber \\
 &  = \sum_\lambda a(\lambda) e^{-\lambda^2(2t+z^2)} \cos(\sqrt{2}\lambda t - \theta_\lambda) \,.        
\end{align}
Here $\lambda$ must be either a real or a pure imaginary number in the following reason. 
If $\lambda$ is a complex number, then the above solution could contain an unstable mode. 
However if such an unstable mode is allowed, it would imply that the Schwarzschild-AdS black brane ($p=0=P$) 
itself would admit an unstable perturbation as we have the same expression (\ref{expr:delta:m}) 
for the perturbations and the same boundary conditions (\ref{condi:z:0}) and (\ref{condi:z:infty}) for the case 
of the Schwarzschild-AdS black brane, which is however thought to be stable from the thermodynamic perspective. 
Now suppose $\lambda$ is pure imaginary. Then $\delta m$ is non-normalizable on 
$t=const.$ surface, hence is not a physically acceptable perturbation. Therefore, $\lambda$ must be a real number, 
for which the perturbation solution~(\ref{expr:delta:m}) exhibits no instability.  
It is thus plausible to argue that our black mushroom should be stable under type of the perturbations considered above. 
This argument is also consistent with the speculation that any black string portion of the neck should be stable 
according to the Gubser-Mitra conjecture~\cite{GubserMitra2000, GubserMitra2001}, as the portion has always positive specific heat.  
To fully justify this stability argument, we however need a thorough study of the dynamical perturbations, which is the near future task.       
%

\section{Spherical black hole case}
\label{sec:V}
In this section, we pay close attention to the polarized AdS black hole with a spherical horizon. 
If such an AdS black hole is highly squashed by external electric field, ``dumbbell" type black hole with a neck 
connecting two spheres appears~(see Fig.~4.). Then, as discussed in the previous sections, Gregory-Laflamme 
instability would occur 
unless the black string portion becomes thermodynamically stable.  
\begin{figure}[htbp]
\centering
  \includegraphics[width=4.5truecm,clip]{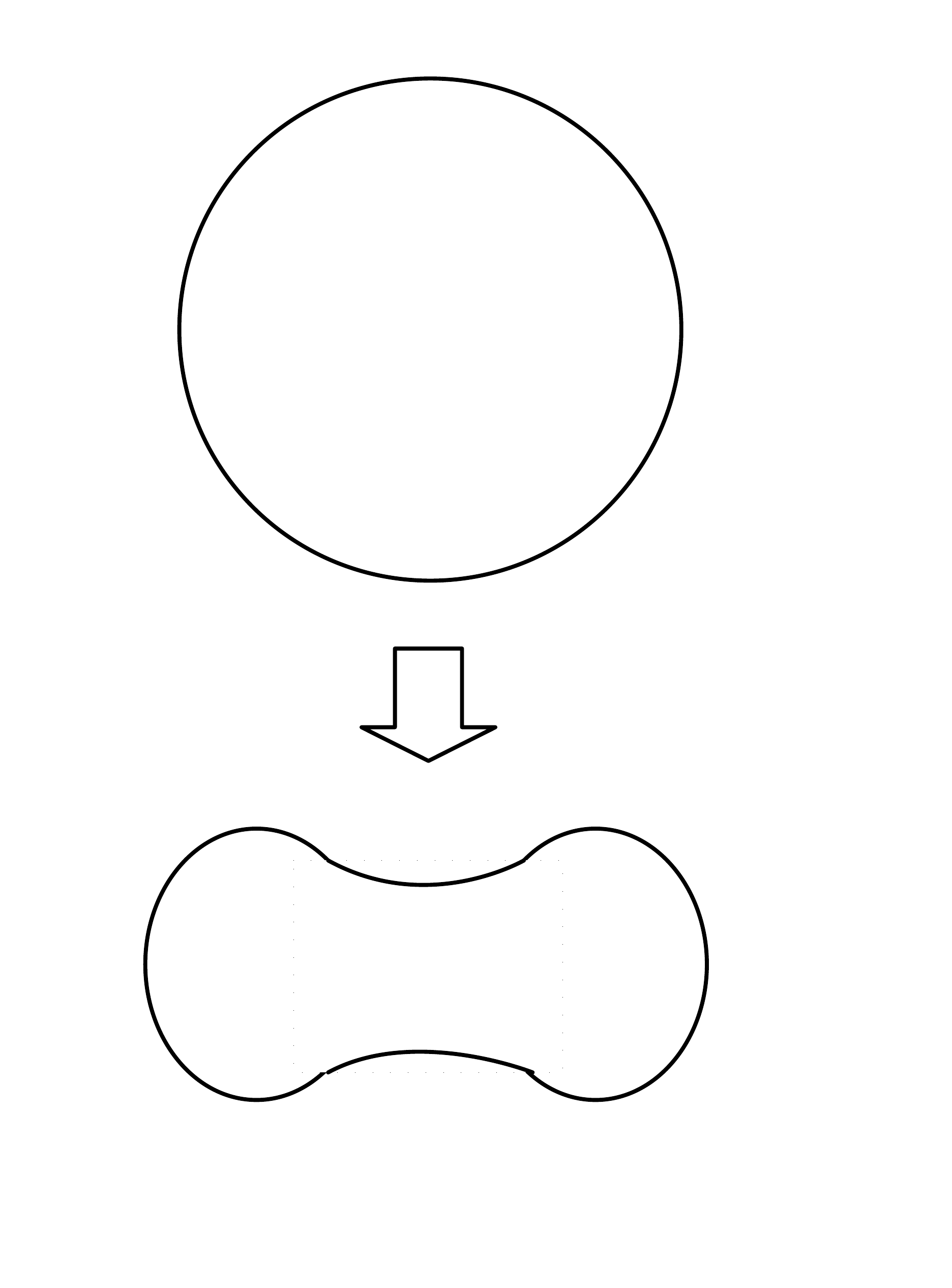} 
\caption{The construction of the ``dumbbell" like black hole solution from the spherically symmetric black hole}
\end{figure}

One might ask whether such a polarized AdS black hole with a spherical horizon is unstable 
or not, because small AdS black holes are thermodynamically unstable. 
In this section, we investigate properties of such a polarized AdS black hole at large $D$ by analyzing 
$1+1$-dimensional effective equations as follows. 

\subsection{Derivation of effective equations}
We make the metric ansatz as 
\begin{align}
\label{metric_ansatz_sp}
& ds^2=-Adt^2+2u_t\,dtdr-2C_z\,dtdz+G_{zz}dz^2 \nonumber \\
& \,\, \, \quad \quad +r^2\sin^2 z\,d\Omega_{n-2}^2 \,, 
\end{align}
where $z$ is the angular coordinate in the range $0\le z\le \pi$. 
As in the brane case, under the condition~(\ref{gauge_ansatz}), the metric is 
expanded as 
\begin{align}
\label{metric_expansion_sp}
A(r,t,z) &=\frac{r^2}{L^2}\left(1-\frac{m(t,z)}{r^n}\right)+1 \nonumber \\
&\quad \,\,\, +\frac{1}{n}\left(\frac{Q(t,z)}{r^{2n-2}}+a_1(r,t,z)\right)+O(n^{-2}), \nonumber \\
 C_z(r,t,z) &= \frac{p(t,z)}{nr^n}+O(n^{-2}), \nonumber \\
 u_t(r,t,z) &=1+\frac{\beta_t(r,\,t,\,z)}{n}+O(n^{-2}), \nonumber \\
 G_{zz} &=r^2+\frac{H(r,\,t,\,z)}{n}+O(n^{-2}), \qquad \nonumber \\ 
 Q & =L^2q^2(t,z) \,. 
\end{align} 
It is easily checked that the metric~(\ref{metric_expansion_sp}) and the gauge fields~(\ref{sol_gauge_t}), 
(\ref{sol_gauge_z}) are the leading order solutions for the spherical case. At next to leading order, 
we can set $\beta_t=H=0$ as in the brane case, and 
we find the solution for $a_1$ as 
\begin{align}
\label{sol_a1_sp}
a_1\simeq-\frac{L^2p\cos z\ln R}{R\sin z} \,. 
\end{align} 
Substituting Eq.~(\ref{sol_a1_sp}) into the Einstein equations~(\ref{Eq:Ein}), 
we obtain evolution equations for $q$, $m$, and $p$ as 
\begin{align}
\label{Eq:q_sp}
&\hspace{-0.1cm}\frac{\p q}{\p t}-\frac{\cos z}{\sin z}\frac{\p q}{\p z}
- \frac{(R_HL^2-m)\cos z}{\sin z}\frac{\p P}{\p z} \nonumber \\
& \qquad +\frac{(L^2+1)\cos z}{m\sin z}pq=0 \,,  \\
\label{Eq:m_sp}
&\hspace{-0.1cm} \frac{\p m}{\p t}+\left((1+L^2)p-\frac{\p m}{\p z} \right)\frac{\cos z}{\sin z}=0 \,,  \\
\label{Eq:p_sp}
&\hspace{-0.1cm}  \frac{\p p}{\p t}+\frac{(1+L^2)p^2\cos z}{m\sin z}- 2q\frac{\p P}{\p z}+\frac{1}{L^2}\frac{\p m}{\p z}
\nonumber \\
& \qquad -2p-\frac{\p }{\p z}\left(\frac{p\cos z}{\sin z} \right)=0 \,,
\end{align}
where $R_H$ is the value of $R$ at the horizon determined by $A=0$. 

\subsection{Properties of static solutions}
Here, we investigate the properties of the static spherical black hole solutions. 
Assuming that $q$, $m$, and $p$ depend on the variable $z$ only, the static equations 
are reduced from Eqs.~(\ref{Eq:q_sp}), (\ref{Eq:m_sp}), and (\ref{Eq:p_sp}) as
\begin{align}
\label{static_eq_sp}
 m' &=(1+L^2)p \,, \nonumber \\
 q' &= \frac{m}{1+L^2}P'+\frac{1+L^2}{m}pq \,, \nonumber \\
 \left(p\cot z\right)' &=\frac{m'}{L^2}+\frac{(1+L^2)p^2}{m}\cot z - 2qP'-2p \,. 
\end{align}
The value of $R$ at the horizon, $R_H$, is determined by $A=0$ in 
Eq.~(\ref{metric_expansion_sp}) as  
\begin{align}
& R_H=\frac{m}{1+L^2} -\frac{1}{n}\left[ -\frac{2mL^2}{(1+L^2)^2}\ln \left(\frac{m}{1+L^2}\right) \right. \nonumber \\
& \quad \qquad \qquad  \qquad \qquad \left. +\frac{L^2q^2}{m}+\frac{ma_1}{(1+L^2)^2} \right]. 
\end{align}
Up to $O(1)$, the temperature $T$ 
is evaluated at the value of surface gravity on the 
horizon, 
\begin{align}
\label{temp_sp}
& \kappa=\frac{1}{2}A_{,r}=\frac{(1+L^2)n}{2L^2}+\frac{1-L^2}{2L^2}\ln \frac{m}{1+L^2} \nonumber \\
&\qquad -\frac{(1+L^2)^2q^2}{2m^2}
-1+\frac{1}{2L^2}-\frac{(1+L^2)p\cos z}{2m\sin z}. 
\end{align}
It is easily checked that $\kappa$ is constant along the horizon by showing $\kappa_{,z}=0$ by 
the static equations~(\ref{static_eq_sp}), as in the black mushroom case. 

Now, we consider the ``dumbbell" type static spherical black hole solutions in which the equatorial plane is squashed by 
the external electric field. For simplicity, we assume that the solution is symmetric with respect to the equatorial plane. 
This means that 
\begin{align}
\label{cond_p_sp}
p|_{z = \frac{\pi}{2}} 
=m' |_{z= \frac{\pi}{2} }=0. 
\end{align}
We also assume that $M$ sufficiently quickly approaches zero at the north pole~(also south pole) as 
\begin{align}
\label{bd_north}
M=cz^{2+\epsilon}, \qquad \epsilon>0, 
\end{align}
as in the black mushroom case. 

The total mass and the charge ${\cal M}$ and ${\cal Q}$ are determined by the mass and charge 
density $m$ and $q$ as   
\begin{align}
\label{total_m}
{\cal M}\sim \int^\pi_0 m(z)\sin^{n-2} z\, dz, \quad {\cal Q}\sim \int^\pi_0 q(z)\sin^{n-2} z\, dz. 
\end{align}
This implies that ${\cal M}$ and ${\cal Q}$ are dominated by the values of $m|_{z=\pi/2}:=m_e$ and $q|_{z=\pi/2}:=q_e$, 
respectively in the large $n$ limit since $\sin^{n-2} z$ becomes zero except $z=\pi/2$ in the limit.  
At the equatorial plane, by Eq.~(\ref{cond_p_sp}), $\kappa$ is rewritten by $m_e$ and $q_e$ as 
\begin{align}
\label{temp_sp1}
& \kappa=\frac{(1+L^2)n}{2L^2}+\frac{1-L^2}{2L^2}\ln \frac{m_e}{1+L^2} \nonumber \\
&\qquad -\frac{(1+L^2)^2q_e^2}{2m_e^2}-1+\frac{1}{2L^2} \,. 
\end{align}
Therefore, the condition for the negative specific heat becomes 
\begin{align}
\label{negative_specific_heat_sp}
L>1  \quad \mbox{and} \quad 2(1+L^2)^2q_e^2<\frac{L^2-1}{L^2}m_e^2 \,.
\end{align}

Let us define $M$ and $\xi$ by Eq.~(\ref{variable_M}). Here, $m_0$ and $q_0$ are defined by the values 
of north pole, respectively: 
\begin{align}
m_0:=m|_{z =0}, \qquad q_0:=q|_{z = 0}. 
\end{align}
Eliminating $p$ from Eqs.~(\ref{static_eq_sp}) and integrating the second equation of (\ref{static_eq_sp}), we find 
\begin{align}
\label{relation_q_P_sp}
q_0= \frac{Pm_0}{1+L^2}, 
\end{align}
where we used the regularity condition $A_t=0$ on the horizon. 
Eliminating $P$ from the third equation in (\ref{static_eq_sp}) by Eq.~(\ref{relation_q_P_sp}), 
we obtain 
\begin{align}
M''\cot z+\left(1-\frac{1}{L^2}-\frac{1}{\sin^2 z}\right)M'=-\frac{2(1+L^2)^2\xi\xi'}{m_0^2}. 
\end{align}
The first integration yields
\begin{align}
 M'\cot z+\left(1-\frac{1}{L^2}\right)M &=\frac{(1+L^2)^2}{m_0^2}(q_0^2-\xi^2) \nonumber \\
&=(1+L^2)^2\left(\frac{q_0^2}{m_0^2}-\frac{q^2}{m^2} \right), 
\end{align} 
where we used $\xi^2/m_0^2=q^2/m^2$ and the boundary condition~(\ref{bd_north}).  
Therefore, we obtain 
\begin{align}
 M|_{z =\frac{\pi}{2}} &=\frac{L^2(1+L^2)^2}{(L^2-1)}\left(\frac{q_0^2}{m_0^2}-\frac{q_e^2}{m_e^2} \right) \nonumber \\
&>-\frac{L^2(1+L^2)^2 }{(L^2-1) } \frac{q_e^2}{m_e^2}>-\frac{1}{2L^2}>-\frac{1}{2}
\end{align}
under the condition~(\ref{negative_specific_heat_sp}). This is the lower bound of $M$ at the equatorial plane, which 
means that the equatorial plane cannot be highly squashed, keeping the negative specific heat. 
In other words, highly squashed equatorial black dumbbell is possible to construct but its specific heat is 
always positive.   
According to the 
Gubser-Mitra conjecture~\cite{GubserMitra2000}, this indicates that Gregory-Laflamme instability does not 
occur in the ``dumbbell" type static spherical black hole solutions. 

\subsection{Linear perturbations}
We consider linear perturbation of the static black hole solutions satisfying Eqs.~(\ref{static_eq_sp}). As in the black 
brane case, we assume that the perturbation of $P$ is zero. Then, linearizing the evolution equations~(\ref{Eq:q_sp}), 
(\ref{Eq:m_sp}), and (\ref{Eq:p_sp}), we obtain     
\begin{align}
\label{pert_q_sp}
& \dot{\delta q}-\left(\delta q' - \frac{P'\delta m}{1+L^2}+\frac{(1+L^2)pq}{m^2}\delta m \right)\cot z
\nonumber \\
& \quad +\frac{1+L^2}{m}(q\delta p+p\delta q)\cot z=0 \,, \\
\label{pert_m_sp}
& \dot{\delta m}+\{(1+L^2)\delta p-\delta m'\}\cot z=0 \,, \\
\label{pert_p_sp}
& \dot{\delta p}+(1+L^2)\left(\frac{2p\delta p}{m}-\frac{p^2}{m^2}\delta m   \right)\cot z
\nonumber \\
&\quad - 2P'\delta q+\frac{\delta m'}{L^2}-2\delta p-(\delta p\cot z)'=0 \,.
\end{align}
From the regularity on the equatorial plane $z=\pi/2$, the following boundary conditions are derived: 
\begin{align}
\dot{\delta q} |_{z = \frac{\pi}{2} }=\dot{\delta m} |_{z = \frac{\pi}{2} }=0 \,.
\end{align}
Note that this is consistent with the mass and charge conservation law, i.~e.~, ${\cal M}$ and 
${\cal Q}$ defined in Eq.~(\ref{total_m}) are constant during the time evolution in the large $n$ limit.

Eliminating $\delta p$ by using Eq.~(\ref{pert_m_sp}), we obtain two equations for $\delta m$ and $\delta q$ as follows:
\begin{align}
&  \dot{\delta q} - \cot z \delta q' + (1+L^2) \frac{p}{m} \cot z \delta q 
\nonumber \\
& \quad - \cot z \left[ (1+L^2)\frac{pq}{m^2} - \frac{P'}{1+L^2} \right] \delta m
\nonumber \\
& \, \quad + (1+L^2)\frac{q}{m}\cot z \delta m' - \frac{q}{m}\dot{\delta m} =0 \,,
\end{align}
\begin{align}
& 
2 \dot{\delta m}' - \frac{1}{\cot z}  \ddot{\delta m}
-\cot z \delta m'' 
\nonumber \\
&\quad + \left[ \cot^2 z + \frac{1}{L^2} \right] \delta m'+ \frac{2}{\cot z} \dot{\delta m} 
 \nonumber \\
& \, \quad + 2(1+L^2)\frac{p}{m}(\cot z \delta m'- \dot{\delta m})
 \nonumber \\ 
&  \,\, \quad -(1+L^2)^2 \cot z \frac{p^2}{m^2} \delta m - 2(1+L^2)P' \delta q = 0 \,.
\end{align}
The stability analysis from now on parallels what we have done below Eqs.~(\ref{eq:m:tz}) and (\ref{eq:q:tz}) for our black branes. 
We can make the same plausible argument for our black dumbbell, agreeing with the thermodynamical argument 
that the Gregory-Laflamme type instability does not occur in the squashed black holes.

\section{Summary and discussions}
\label{sec:VI}
In this paper, we have first derived a new set of effective equations \eqref{Eq_mass} - \eqref{eq_q}, describing the dynamics of the polarized black branes (holes) 
to leading order in the $1/D$ expansion and using these, 
we have tested cosmic censorship conjecture in polarized AdS black brane~(hole) solutions at large $D$ 
dimensions. As expected in the four-dimensional analysis~\cite{HSW2016}, we found a black mushroom solution 
where a black hole is connected with an asymptotic planar black brane through a black-string-shape neck under the 
localized chemical potential. Contrary to our first naive expectation, the black-string-shape neck part is thermodynamically stable. 
This indicates that the localized string cannot be pinched off dynamically according to the Gubser-Mitra 
conjecture~\cite{GubserMitra2000, GubserMitra2001}. 
We have extended the analysis to the AdS black hole case and found that highly squashed black hole is also 
dynamically and thermodynamically stable. 
These facts imply that the cosmic censorship is not violated in such polarized AdS black brane~(hole) solutions 
at large $D$ by the Gregory-Laflamme instability~\cite{GL}. 
  
 
For simplicity, we have treated the gauge field as a probe approximation in the sense that the horizon geometry at 
leading order is neutral black brane~(hole) solutions. In other words, the horizon is embedded 
at the fixed bulk radial  
coordinate in AdS spacetime in the leading order. 
To take into account the gauge field at leading order, we must construct a charged polarized black brane~(hole) solutions 
at leading order so that the horizon is located over different radial region. It is interesting to test the cosmic censorship 
conjecture in that case. This will be investigated in the near future.       

\bigskip

\noindent 
{\bf Acknowledgments} 
We would like to thank Kentaro Tanabe and Norihiro Tanahashi 
for discussions in the early stage 
of the project. We would especially like to thank Kentaro Tanabe for sharing his unpublished notes  \cite{tanabenote} with us in the early stage of the project, 
and Roberto Emparan for valuable comments on various aspects of our results.    
We would also like to thank Gary T.~Horowitz and Ryotaku Suzuki for useful comments on the manuscript. 
This work was supported in part by JSPS KAKENHI Grant Number 25800143 (NI), 15K05092 (AI), 17K05451 (KM).
 
\appendix
\section{Formula for curvature decomposition}
$D$-dimensional Ricci curvature on the metric ansatz~(\ref{metric_ansatz_br}) is decomposed into Ricci curvature and 
the Christoffel symbol on the three-dimensional spacetime $(t,\,r,\,z)$ as 
\begin{align}
\label{decomposition}
& R_{rr}={R^{(3)}}_{rr}+(n-2)\left(\frac{{\Gamma^r}_{rr}}{r}+\frac{{\Gamma^z}_{rr}}{z}   \right), \nonumber \\
& R_{rz}={R^{(3)}}_{rz}+(n-2)\left(\frac{{\Gamma^z}_{zr}}{z}+\frac{{\Gamma^r}_{rz}}{r}-\frac{1}{rz}   \right), \nonumber \\
& R_{rt}={R^{(3)}}_{rt}+(n-2)\left(\frac{{\Gamma^z}_{rt}}{z}+\frac{{\Gamma^r}_{rt}}{r}   \right), \nonumber \\
& R_{tt}={R^{(3)}}_{tt}+(n-2)\left(\frac{{\Gamma^z}_{tt}}{z}+\frac{{\Gamma^r}_{tt}}{r}   \right), \nonumber \\
& R_{zz}={R^{(3)}}_{zz}+(n-2)\left(\frac{{\Gamma^z}_{zz}}{z}+\frac{{\Gamma^r}_{zz}}{r}   \right), \nonumber \\
& R_{tz}={R^{(3)}}_{tz}+(n-2)\left(\frac{{\Gamma^z}_{tz}}{z}+\frac{{\Gamma^r}_{tz}}{r}   \right). 
\end{align}


\end{document}